# The Distinction Between Thermal Nonequilibrium and Thermal Instability

James A. Klimchuk

Heliophysics Science Division
NASA Goddard Space Flight Center
Greenbelt, MD  20771  USA
James.A.Klimchuk@nasa.gov

**Abstract**

For some forms of steady heating, coronal loops are in a state of thermal nonequilibrium and evolve in a manner that includes accelerated cooling, often resulting in the formation of a cold condensation. This is frequently confused with thermal instability, but the two are in fact fundamentally different. We explain the distinction and discuss situations where they may be interconnected. Large-amplitude perturbations, perhaps associated with MHD waves, likely play a role in explaining phenomena that have been attributed to thermal nonequilibrium but also seem to require cross-field communication.

## 1.  Thermal Nonequilibrium

Thermal nonequilibrium (TNE) is a fascinating property that may explain a number of solar phenomena, including coronal rain, prominence formation, long-period loop pulsations, and quasi-periodic disturbances in the solar wind. It has been suggested that TNE may be even more widespread than these relatively isolated occurrences, perhaps involving a sizable fraction of the magnetic flux in an active region. Although the existence of TNE was noted by Serio et al. (1981), it was first studied in detail by Antiochos and Klimchuk (1991). There has been considerable work on the topic since then, especially recently, and the reader is referred to the introductions by Froment et al. (2018), Klimchuk and Luna (2019), and Antolin (2019) for more information and references.

Thermal nonequilibrium is often confused with thermal instability. The two share common properties, but they are fundamentally different. The purpose of this article is to highlight the differences and to discuss situations where they may be interconnected.

Thermal nonequilibrium describes a state of the plasma contained in a magnetic flux tube that is rooted to the solar surface at both ends, *i.e.*, a coronal loop. The loop can be an observationally distinct feature in an image or an indistinguishable part of the diffuse emission. In the following discussion, we assume that the magnetic field is rigid and that any evolution involves the plasma only. Possible MHD effects are mentioned at the end.

Theoretical modeling reveals that steady coronal heating usually results in an equilibrium, where the plasma does not evolve. If the heating and magnetic geometry are symmetric, it is a static equilibrium, and the plasma is at rest. Asymmetries lead to a modified equilibrium that includes a steady end-to-end flow. This is sometimes called a syphon flow, because it is driven by a pressure differential between the two sides.

Something very interesting can occur when steady heating is concentrated in the low corona. If it decreases sufficiently rapidly with distance from the loop footpoints, no equilibrium is possible. No combination of temperatures, densities, and velocities is able to produce a steady state. The loop continually evolves even though the heating is constant in time. Essentially, the loop is searching for a nonexistent equilibrium. The conditions required for TNE are discussed by Klimchuk and Luna (2019). As a rough rule of thumb, the ratio of apex to footpoint heating rates must be less than about 0.1, and asymmetries in heating and/or cross-sectional area must be less than about a factor of 3.

Loops in TNE undergo a characteristic cyclical evolution, with wide swings in temperature and density. Throughout most of the cycle, the loop broadly resembles a "typical" coronal loop (Reale, 2014), with temperature of order $10^6$ K and density of order $10^9$ cm$^{-3}$. Compared to equilibrium conditions, however, the plasma is under-dense in the lower legs and over-dense in the upper section. The under-density leads to an evaporation of plasma from the footpoints as the legs attempt to establish a local equilibrium. The evaporated plasma is of course not confined to the legs and fills the entire loop, exacerbating the over-dense conditions at the top. Higher densities cause stronger radiation, and the loop cools, slowly at first, but at an ever increasing rate. This usually culminates in the formation of a cold (chromospheric temperature) high-density condensation. Pressure gradients cannot support the excess weight, and the condensation slides down a leg to the solar surface, evacuating the loop in the process. Because the heating is unchanged, the low-density plasma that remains is quickly heated to high temperatures, and a new cycle begins. In some circumstances, the plasma reheats before the temperature drops below 1MK. This is known as an incomplete condensation and is not yet understood (Mikic et al., 2013).

Thermal nonequilibrium is an example of a limit cycle, in which the system (loop) retraces the same path within phase space (e.g., temperature versus density). The possibility of limit cycles in coronal loops was first pointed out by Kuin and Martens (1982), who emphasized the importance of the mass exchange between the corona and chromosphere. Using a simplified model, they found limit cycles even in loops with uniform heating. We now know from more sophisticated modeling that this type of behavior is only possible with highly stratified heating.

The number of repeating TNE cycles that a loop experiences depends on how long the TNE conditions are maintained. Cycle periods in simulations generally range between 2 and 15 hours, depending on the details of the heating profile and loop geometry, including heating magnitude and stratification, loop length, cross-sectional area variation, and shape (Froment et al., 2018; Winebarger et al., 2018; Müller, Hansteen, and Hardi, 2003). Since heating and geometry can evolve over timescales of hours or less on the Sun, irregular cycles and restricted numbers of cycles, including just a single cycle, are not unexpected. For example, coronal rain typically does not repeat in a regular manner (P. Antolin and C. Froment, private communication), though examples of periodic rain have been reported (Auchère et al. 2018). Further study is needed for proper statistics.

Prominences represent a somewhat different situation. If the condensation settles into a dip in the magnetic field, there is just one cycle even if the heating and geometry remain steady indefinitely. The presence of the stationary condensation in the dip allows the loop to establish a quasi-equilibrium that was not possible before the condensation formed. We can think of the evolved system as being comprised of two separate halves, each extending from a footpoint in the chromosphere to a "footpoint" at the condensation. A quasi-steady upflow occurs because the heating is much stronger at one end (near the chromosphere) than at the other end (near the condensation). Material evaporates from the chromosphere and condenses onto the slowly growing prominence mass from both sides. The sequence of evolution is thus: TNE conditions in the full loop → condensation formation in the magnetic dip → steady flow conditions in the two "half loops." There is a single condensation cycle even though the heating and geometry never change.

## 2. Thermal Instability

Since the cooling that occurs during the first phase of a TNE cycle proceeds at an ever faster rate, it resembles a thermal instability. Many people refer to it as such, but this is incorrect. The McGraw-Hill Dictionary of Scientific and Technical Terms defines instability to be "a property of the steady state of a system such that certain disturbances or perturbations introduced into the steady state will increase in magnitude, the maximum perturbation amplitude always remaining larger than the initial amplitude" (Parker, 1994). With TNE, there is no equilibrium (no steady state), so it is meaningless to talk about its stability. The system cannot return to its pre-perturbation state (in a given cycle) even if the perturbation shrinks, because the system is inherently time dependent. We recommend that the cooling that occurs during a TNE cycle be referred to as a *thermal runaway* to distinguish it from thermal instability.

We note that limit cycles are sometimes described as stable or unstable. In a stable limit cycle, the perturbed system eventually returns to the same closed path in phase space. In an unstable cycle, it does not. Simulations show that TNE generally corresponds to stable limit cycles, though irregular/unstable cycles are also seen (Müller, Hansteen, and Peter, 2003; Müller, Peter, and Hansteen, 2004; Müller et al., 2005). The thermal instability we discuss in this article is different from the instability of limit cycles. The standard definition requires the existence of an equilibrium, which of course is not the case in limit cycles.

Uniform coronal plasmas are in equilibrium if the temperature and density provide a perfect balance between radiation and heating. Such equilibria are well-known to be thermally unstable (Parker, 1953; Field, 1965). Because the radiative loss function varies inversely with temperature-

--$\Lambda(T) \propto T^b$, where $b < 0$---a decrease in temperature causes stronger radiation, which decreases the temperature further, and a runaway cooling ensues.

If the temperature perturbation is local rather than uniform, then thermal conduction must be taken into account. Energy is conducted into the local temperature depression, and this has a stabilizing effect. Since the magnitude of the conduction flux is proportional to the temperature gradient, the stabilization is weaker for long-wavelength perturbations than for short-wavelength perturbations. Beyond a certain length, conductive heating is unable to compensate for the enhanced radiative cooling, and the perturbation grows. Uniform flux tubes that are subjected to a broad spectrum of perturbations are therefore thermally unstable.

A crucial point that is often overlooked is that coronal loops are *not* uniform flux tubes. Although temperature and density vary slowly with position along most of the loop, the coefficient of thermal conduction, $\kappa \propto T^{5/2}$, is so large at coronal temperatures that a small gradient carries a large energy flux. The divergence of that flux is an important term in the energy balance. In fact, cooling from thermal conduction exceeds cooling from radiation by a factor of 2-4 in the coronae of equilibrium loops with uniform heating (Klimchuk, Patsourakos, and Cargill, 2008; Cargill, Bradshaw, and Klimchuk, 2012).

A strong downward conduction flux from the corona is primarily what powers the intense radiation from the transition region. The corona, transition region, and chromosphere are thus fundamentally coupled---energetically and dynamically. A change in one part of the loop affects the entire loop. As a consequence, most loops are thermally stable to even long-wavelength perturbations (Klimchuk, Antiochos, and Mariska, 1987). Equilibrium loops that are on the verge of TNE may be an exception, as we discuss below. The stability of equilibrium loops is demonstrated by the many numerical studies in which an initial state is relaxed to an equilibrium in the presence of steady uniform heating, after which some additional type of heating is imposed. If the equilibria were unstable, this procedure would fail. Examples of studies with many loops of variable length and heating rate include Warren and Winebarger (2007) and Luna, Karpen, and DeVore (2012). No difficulties in relaxing to an equilibrium were encountered (H. Warren and M. Luna, private communication).

The above discussion concerns equilibria with uniform heating or heating with a modest spatial dependence. As the heating becomes more and more concentrated at low altitudes, the temperature profile of the loop (temperature versus position along the loop axis) becomes less and less rounded. For strong enough heating concentration, the profile is nearly flat in the central section that spans the apex. Thermal conduction is much less important in the energy balance under these conditions, and it is reasonable to imagine that these loops, which are on the verge of TNE, may be thermally unstable.

It must be remembered that realistic loops are not perfectly symmetric and therefore contain flows. Depending on the degree of asymmetry, the enthalpy flux carried by the flow can be energetically important. How this affects stability has yet to be determined. A related question concerns the growth rate of the instability. If a perturbation grows, does it have time to reach substantial amplitude before it is carried to the chromosphere by the flow?

As we have discussed, strong heating stratification leads to TNE. If the stratification is strong, but not too strong, it is possible to have equilibria with small temperature inversions, *i.e.*, shallow dips in the temperature profile (Winebarger, Warren, and Mariska, 2003; Müller, Peter, and Hansteen, 2004; Müller et al., 2005; Martens, 2010). An interesting question is whether equilibria with deep dips are also possible. They are not seen in relaxed solutions to the time-dependent equations, but that could be because they are unstable. We show in the appendix that such equilibria are in fact unlikely.

### 3. Growing Perturbations During a TNE Cycle

We have argued that the formation of a condensation during a TNE cycle cannot be formally classified as a thermal instability because there is no equilibrium to go unstable. However, if the timescale for evolution in the absence of a perturbation is much longer than the perturbation growth time, then it is physically reasonable to treat it like a thermal instability. A classical perturbation analysis would be meaningful. This is the view adopted by Xia et al. (2011), Claes and Keppens (2019), and Antolin (2019). We support this approach, as long as all of the important physical effects are included and as long as it is demonstrated that the system would evolve significantly more slowly without a perturbation than with one. To our knowledge, these requirements have not yet been met.

If equilibria having nearly flat temperature profiles are prone to instability---yet to be established---what about TNE loops at the time during their cycle when the temperature profile is approximately flat? The situations are similar, but there is one major difference. TNE loops have significant evaporative upflows in both legs, and they are energetically important. For example, at the time when the temperature profile is flat in the base model of Klimchuk and Luna (2019, black curve in Fig. 6), the heating from the divergence of the enthalpy flux is nearly equal to radiative cooling, with coronal heating being much smaller than either (at the apex). Whether the flows are stabilizing or destabilizing has yet to be determined.

It has been suggested that condensations such as coronal rain must be caused by thermal instability and would not come about from TNE alone (Antolin, 2019). The reasoning is that TNE is a global phenomenon, involving the entire loop, whereas condensations are small features that require a localized effect. The weakness of this argument is that radiative cooling is not a linear process. Cooler plasmas cool at a faster rate than warmer plasmas, so temperature profiles do not maintain their shape. A broad shallow dip becomes progressively more narrow as it deepens.

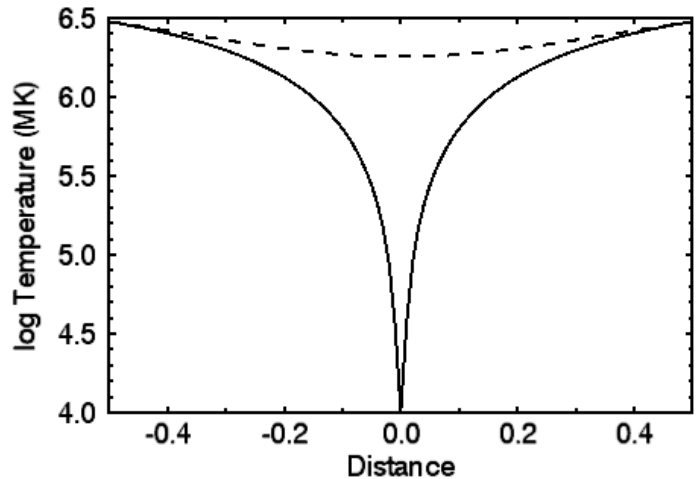

**Figure 1** Temperature versus position along a "loop" that begins slightly out of equilibrium: initial profile (dashed) with the dip magnified by a factor of 40, and final profile (solid). See text for details.

Figure 1 shows two temperature profiles from a highly simplified yet instructive model. We begin with a coronal plasma having a uniform density of $5 \times 10^9$ cm$^{-3}$ and a nearly uniform temperature of $3 \times 10^6$ K. The otherwise flat temperature profile is modified by a half cosine wave with an amplitude of $3 \times 10^4$ K, or 1%. It has been magnified by a factor of 40 in the figure (dashed curve) in order to be distinguishable from a straight line. We impose a constant uniform heating that would maintain the plasma at $3 \times 10^6$ K if the dip were not present. The model is meant to be an idealized representation of the central section of a TNE loop at a time just after the temperature inversion first appears.

We adopt a power law radiative loss function of slope b = -0.5. Because the temperature is less than $3 \times 10^6$ K, there is a slight excess in radiation compared to heating. We allow each element of plasma to cool at constant density. Thermal conduction and flows are ignored. The cooler plasma at the center of the dip radiates slightly more strongly than the rest. The difference is very small at first, but increases rapidly as the temperatures separate. This causes the temperature profile to narrow dramatically as it deepens. The solid curve shows the profile at the time chromospheric temperatures are first reached. Although this model is highly idealized, it demonstrates vividly how small condensations can form, without instability, from a coronal system that is globally out of equilibrium. This is exactly the situation in TNE.

The same model can also represent a different situation---that of a true equilibrium that has been perturbed by a 1% long-wavelength disturbance. The evolution is identical. This emphasizes a crucial point: the physics that governs the thermal runaway in a TNE loop is equivalent to the physics that governs the thermal runaway in an unstable equilibrium loop. In one case the loop is inherently out of equilibrium, and in the other it is forced out of equilibrium by an applied perturbation. The timescale of the evolution is similar in the two cases whenever the perturbation creates a similar imbalance between heating and cooling. Note that a short-wavelength perturbation would grow more slowly than a long-wavelength perturbation due to the stabilizing effect of thermal conduction (not included in the simple model).

We have demonstrated that the formation of condensations does not require thermal instability, but we do not wish to imply that growing perturbations cannot be important. On the contrary, we argue below that they may be crucial in some situations. A key point is that, in order for a perturbation to drive the evolution of a TNE loop and influence the final outcome, it must produce a significantly larger deviation from equilibrium than was already present. This suggests that the initial amplitude must be large.

## 4. Cross-Field Communication

Several observed phenomena believed to involve TNE are difficult to explain without some type of cross-field communication. Coronal rain sometimes appears in "showers" whereby several condensations form at roughly the same time (Antolin and Rouppe van der Voort, 2012). Individual condensations sometimes have a large cross-field extent, spanning many adjacent loops (F. Auchère, private communication). Long-period pulsating loops appear in groups and repeat, in phase, with a regular period over many cycles (Auchère et al., 2014, 2018; Froment et al., 2015).

It is rather difficult to imagine that highly similar TNE conditions would turn on simultaneously in multiple loops to produce the collective behavior demonstrated by these phenomena. The heating rate, heating profile, and magnetic geometry, including cross-sectional area variation, would all need to be remarkably similar. Pulsating loops present a special challenge. In order for the TNE cycles to remain in phase for many cycles, the conditions would not only need to be essentially identical in the different loops, but they would need to remain so for several days. It seems more plausible that these phenomena involve a cross-field communication, sometimes referred to as "sympathetic cooling" (Antolin, 2019). MHD waves are a likely mechanism. Simulations indicating this type of behavior have been performed (Fang et al., 2015; Xia, Keppens, and Fang, 2017; Claes and Keppens, 2019), but further work is required. Whether the waves come from a single source, or whether a thermal runaway in one loop triggers an adjacent loop, which triggers the next loop, etc. is unclear. It would seem that loops which are close to the boundary between TNE and equilibrium are most susceptible to this type of triggering, since the requirement on the perturbation amplitude is less severe.

## 5. Summary

In summary, thermal instability and thermal nonequilibrium are fundamentally different, even though they both produce a thermal runaway that is driven by the same physics---an accelerating imbalance between heating and cooling. Thermal instability requires an equilibrium which goes unstable when perturbed, even by a small perturbation, while TNE is a situation in which no equilibrium exists. Cold condensations, such as coronal rain, can be produced by TNE alone. Thermal instability is not required. Perturbations may nonetheless be important if they are of large enough amplitude to substantially increase the deviation from equilibrium, *i.e.*, the heating/cooling imbalance. Phenomena like long-period pulsating loops that stay in phase for many cycles seem to involve a cross-field communication that is perhaps best explained by large-amplitude perturbations associated with MHD waves. Further research on the response to perturbations using both MHD simulations and one-dimensional hydrodynamic (loop) simulations should shed new light on this fascinating topic.

Acknowledgments

We acknowledge useful discussions with many individuals, but especially members of International Space Science Institute team on "Observed Multi-Scale Variability of Coronal Loops as a Probe of Coronal Heating," led by C. Froment and P. Antolin. The article was written at the request of several team members. It was supported by the Internal Scientist Funding Model (competitive work package) program at the Goddard Space Flight Center.

Disclosure of Conflicts of Interest:  The author declares that he has no conflict of interest.

**Appendix**

Equilibria with shallow temperature dips are known to exist (Winebarger, Warren, and Mariska 2003; Müller, Peter, and Hansteen 2004; Müller et al. 2005; Martens 2010), and we here consider whether solutions with deep dips are also possible. There is a fundamental difficulty in satisfying energy balance everywhere along a loop with a deep temperature dip. A large heating stratification is necessary to produce the dip, so the heating rate is very small at the apex compared to the base. The density throughout the loop, including the apex, is mostly set by the base heating rate (Klimchuk and Luna 2019), and since the base heating rate is large, so too is the apex density. Radiative losses increase with density and decrease with temperature, and therefore the only way to keep the losses small at the apex, to match the small heating rate, is to have a relatively high temperature. The dip cannot be deep.

Thermal conduction helps mitigate the problem by pumping energy into the dip, allowing a lower temperature, but there is a limit to its effectiveness. To have a sizable thermal conduction flux at low temperatures requires a steep temperature gradient. This generally implies a footpoint-like transition region, with temperatures extending all the way down to chromosphere values. In other words, a cold condensation is unavoidable part of the solution once the dip exceeds some threshhold. This is not a static solution, but one with steady evaporative upflows, as discussed in Section 2 in the context of prominences.

We now provide a more quantitative argument. Consider an equilibrium with a dip, and assume symmetry so that flows can be ignored. There are two locations in each half of the loop where the divergence of the thermal conduction flux vanishes: one at the top of the transition region and another roughly midway down the dip. We use the subscripts "tr" and "d" to designate these locations. The energy balance at both locations is between heating and radiation:

$$Q = \Lambda_0 n^2 T^b,$$

(1)

where $Q$ is the volumetric heating rate and $n$ is the electron number density. Using the ideal gas law and assuming that the gravitation scale height is significantly larger than the geometric height so that pressure is approximately uniform, we have

$$\frac{T_d}{T_{tr}} = \left(\frac{Q_{tr}}{Q_d}\right)^{\frac{1}{2-b}}.$$

(2)

The heating requirements for a temperature inversion are similar to those for TNE: the base heating rate must exceed approximately 10 times the apex heating rate (Klimchuk and Luna 2019). With $Q_{tr} > 10 Q_d$ and $b = -1/2$, Equation 2 implies that $T_d > 2.5 T_{tr}$. Since $T_{tr}$ is typically about 60% of the maximum temperature in the loop (Klimchuk, Patsourakos and Cargill 2008; Cargill, Bradshaw, and Klimchuk 2012), the temperature in the dip would exceed the maximum temperature, which of course is nonsensical. Apparently, for a limited range of heating stratifications, the loop is able to adjust the temperature profile so that $Q_{tr}/Q_d$ satisfies the conditions for a shallow dip. Deep dips would seem to be impossible.